\documentclass{elsart}
\usepackage{graphicx}
\usepackage{url}

\begin{document}
\begin{flushright}
Report NPI ASCR \v Re\v z\\
TECH--12/2008
\end{flushright}
\begin{frontmatter}
\title{Optimal distribution of measurement time in single channel measurements}
\author{J.\,Ka\v spar\corauthref{tel}},
\ead{kaspar@ujf.cas.cz}
\corauth[tel]{tel: +420\,212\,241\,677, fax: +420\,910\,256\,689}
\author{M. Ry\v sav\' y}
\ead{rysavy@ujf.cas.cz}
\address{Nuclear Physics Institute, ASCR\\
CZ--250 68 Rez near Prague, Czech Republic}

\begin{abstract}
Single channel measurements play a minor role in today physics, but they are sometimes unavoidable. Comparing to multichannel measurements, there is distribution of measurement time to be chosen in an experiment design. A method to optimize distribution of measurement time is given, where optimal distribution minimizes standard deviation of a selected fit parameter. As an example, the method is applied to electron spectroscopy experiments.

\end{abstract}
\end{frontmatter}

\section{Introduction}
\thispagestyle{plain}

Single channel measurements offer a unique challenge. Regarding a setup of such experiment, there is one more characteristics to be fixed in advance during the experiment setup, compared to multichannel measurements. This is the distribution of measurement time.

Let there be a single channel experiment in its design phase. A spectrum to be recorded by the experiment is dependent on several parameters. Experimenters are often interested in one of them. Let the mechanical design of experiment be fixed, and let the measurement points be chosen. So, it is known how to measure, and where to measure. However, it is unknown for how long to measure in the particular measurement points, which we call a distribution of measurement time.

Then the task is to distribute the measurement time into the measurement points in the way, that minimizes the standard deviation of the parameter, in which the experimenters are interested. Up to our knowledge, there are no studies available concerning this aspect. The point is, we can not set the measurement times for all the points simultaneously (it is numerically not feasible to minimize a function of tens or hundreds of parameters). Instead, we can fix the measurement time in points one by one, approaching the optimal distribution of measurement time in an iterative way.

\section{Method}

Let $\{E_i\}$, $i = 1\,...\,n$ be a fixed set of measurement points, where $n$ is their number and $\tau_{\rm tot}$ is the total measurement time. Further, let $\{T_i\}$ be the initial time distribution chosen intuitively, e.g. uniform distribution, meaning that equal time is spent in each point. Let the initial time distribution result in an initial standard deviation of the selected fit parameter $\sigma_{\rm init} = \sigma(\{T_i\})$ that was obtained by simulations (creation of pseudo-experimental spectra and their evaluation).

Now, we minimize standard deviation of the selected parameter varying the time $T_1$ in the first measurement point only.
(This means that we simulate spectra corresponding to various $T_1$ and evaluate them to find the value of $T_1$ supplying the minimum standard deviation of the selected fit parameter $\sigma$.
The measurement times in all the rest points are scaled by the same factor to keep the total measurement time equal to $\tau_{\rm tot}$. As the result we obtain the time distribution $\{T^{(1)}_i\}$ and the corresponding standard deviation of the selected fit parameter $\sigma^{(1)} = \sigma(\{T^{(1)}_i\})$. So, the measurement time in the first point is $T^{(1)}_1$, and the measurement time in each of the rest points is equal to
\begin{equation}
	T^{(1)}_i = T_i \times \frac{\tau_{\rm tot} - T^{(1)}_1}{\tau_{\rm tot} - T_1} \quad ,
\label{eqn:sc_factor}
\end{equation}
$i = 2\,...\,n$. Then we do the same for all the other measurement points (starting always with the $\{T_i\}$ distribution). So, we get $n$ time distributions $\{T^{(k)}_i\}\, , \, k = 1\, ...\, n$ and corresponding sigmas $\sigma^{(k)}$.
(For the sake of clarity, we repeat that $\sigma^{(k)}$ are standard deviations of the {\it same} parameter but obtained from spectra corresponding to different time distributions $T_i^{(k)}$.)

Now we construct the best final time distribution from the set $\{T^{(k)}_i\}$ of the partial ones. There are several possibilities how to do it. Experimentally, it turned out that a good way is a weighted sum
\begin{equation}
T'_i = {1 \over n} \sum_{k=1}^n \omega_k T^{(k)}_i \quad ,
\label{eqn:time_dist}
\end{equation}

The weight factors $\omega_k$ may be estimated as:
\begin{equation}
\omega_k = \left( {\sigma_{\rm init} - \sigma^{(k)} \over {1\over n} \sum \sigma^{(k)}}\right)^s \quad ,
\label{eqn:weight_fact}
\end{equation}

where $s$ is chosen to minimize $\sigma(\{T'_i\})$, i.e., we simulate spectra corresponding to various $s$ and evaluate them to find the value of $s$ minimizing the standard deviation $\sigma$.

Having the new time distribution $\{T'_i\}$ we renormalize it with respect to the $\tau_{\rm tot}$. (Note, that the weight factors in eq. (\ref{eqn:weight_fact}) do not keep the $\sum T'_i = \tau_{\rm tot}$.)
Then, we use it instead to the original $\{T_i\}$ to repeat the whole process in an iterative way.

Based on our extensive simulations, this seems to be the best way how to combine the partial time distributions $\{T^{(k)}_i\}$ together, guaranteeing convergence, numerical stability, and reasonable speed. The partial time distributions that significantly improved the standard deviation of the selected parameter are favored, and the convergence speed is enhanced by the choice of the $s$ factor. (Note, that eq. (\ref{eqn:weight_fact}) is well defined because $\sigma_{\rm init} \geq \sigma^{(k)}$ for any $k$.)

We also tried several other ways how to merge the partial time distributions $\{T^{(k)}_i\}$ together. In particular, following choice of the final time distribution:
\begin{equation}
T'_i = T_i^{(i)} \times \frac{\sum_i T_i}{\sum_i T_i^{(i)}}
\label{eqn:time_dist2}
\end{equation}

looked feasible and promised fast convergence to the optimal time distribution. Nevertheless, it proved to be highly numerically unstable, and it oscillated practically always. Finally, if the numerical stability is favored, we tested that setting all the $\omega_k$ in eq. (\ref{eqn:time_dist}) equally to one was a good choice. However, the convergence is about 5 times slower. Practically, we implement the method using a simplex minimizer provided by Minuit2 library wrapped inside the ROOT framework \cite{root}.

\section{Examples}

\begin{figure}
	\begin{center}
		\includegraphics{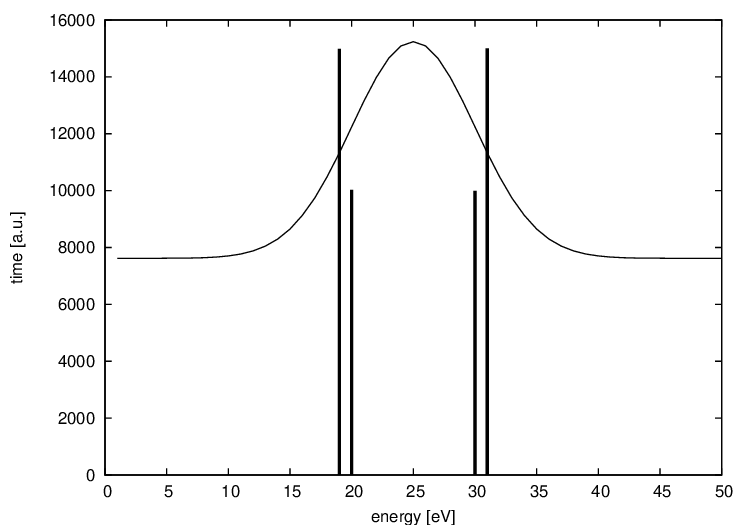}
	\end{center}
	\caption{An example of the optimal distribution of measurement time. The time distribution minimizes the standard deviation of Gaussian line position fitting four parameters: amplitude, background, position, and width. The initial parameter values are given in text. For convenience the Gaussian line is shown as well.}
	\label{fig:pos}
\end{figure}

To test the method we have chosen an idealistic Gaussian line on a constant background as measured by a differential spectrometer:
\begin{equation}
G(E|A,B,E_0,\sigma) = A \cdot \exp\left(\frac{-(E - E_0)^2}{2\sigma^2}\right) + B \quad .
\label{eqn:eqn_gauss}
\end{equation}

The line is described by four fit parameters: amplitude $A$, background $B$, position $E_0$ and width $\sigma$. The initial parameter values were chosen as follows: the amplitude of 10 Hz, the background of 10 Hz, the line position of 25 eV, and the width (sigma) of 5 eV. The distribution of measurement points was chosen to be an uniform one from 0 eV up to 50 eV with a step of 1 eV.

\begin{figure}
	\begin{center}
		\includegraphics{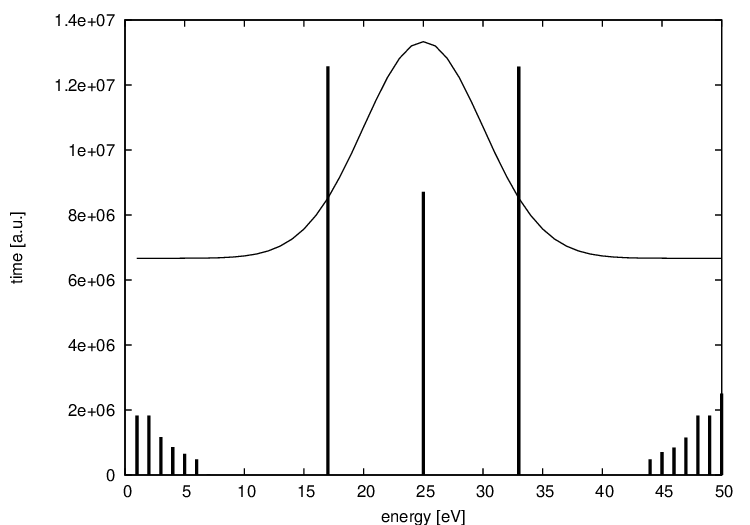}
	\end{center}
	\caption{An example of the optimal distribution of measurement time. The time distribution minimizes the standard deviation of Gaussian line width fitting four parameters: amplitude, background, position, and width. The initial parameter values are given in text. For convenience the Gaussian line is shown as well.}
	\label{fig:sig}
\end{figure}

We studied two cases; optimal time distribution with respect to the standard deviation of the Gaussian line position and that with respect to the line width.
As for the line {\it position}, the resulting optimum time distribution is given in fig. \ref{fig:pos}. It turned out, surprisingly,
that the most of the relevant information for this case is concentrated in four points only. The standard deviation of the line position improved by a factor of 2.0 with respect to the uniform time distribution.
When optimizing the line {\it width} standard deviation, we reached the time distribution depicted in fig. \ref{fig:sig}. Here, the particular standard deviation improved by a factor of 1.4. (In both cases, the optimization procedure needed 100 iterations.)

As the second example, we have chosen an electron spectroscopy experiment to determine the neutrino mass, in particular, the KATRIN experiment \cite{katrin_loi, katrin_des} aiming at neutrino mass sensitivity of $0.2\,$eV/$c^2$. The necessity of the single channel measurement originates from the intrinsic property of the applied spectrometer type, that is the only one exhibiting simultaneously required energy resolution and luminosity. There are four fit parameters in the experiment: the neutrino mass, beta spectrum endpoint, amplitude and background.
Integrated beta spectrum as proposed to be measured by the KATRIN experiment is given in \cite{katrin_des}, as well as the standard distribution of the measurement points, which is a non-equidistant one.

\begin{figure}
	\begin{center}
		\includegraphics{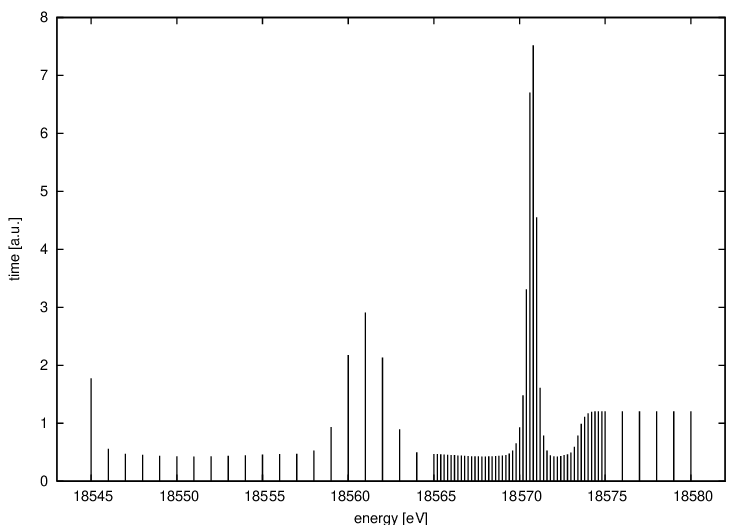}
	\end{center}
	\caption{An example of the optimal distribution of measurement time, which could be used by the KATRIN experiment. The time distribution minimizes the standard deviation of the neutrino mass when fitting four parameters: amplitude, background, the tritium $\beta$-spectrum endpoint, and the neutrino mass. Note, the initial free neutrino mass of 0 eV, and the initial free endpoint of 18\,575 eV was chosen to run the simulations.}
	\label{fig:mnu}
\end{figure}

First, let the neutrino mass be the selected parameter. Then, starting with the uniform time distribution, and optimizing the time distribution in 16 iterations, the simulation resulted in the optimal time distribution shown in fig. \ref{fig:mnu}. The standard deviation of the neutrino mass was improved by factor $1.18$ compared to the uniform time distribution. The same improvement could be achieved by a prolongation of the total measurement time by factor of $1.9$. (Standard deviation of the neutrino mass scales with fourth root of the total measurement time, since the neutrino mass {\it squared} enters the form of tritium beta spectrum.)

\begin{figure}
	\begin{center}
		\includegraphics{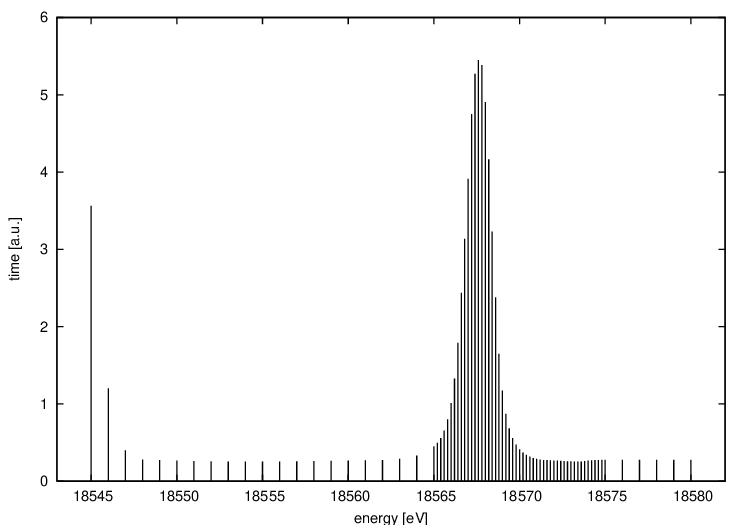}
	\end{center}
	\caption{An optimal time distribution with respect to the minimal standard deviation of the endpoint energy assuming a fixed and known neutrino mass, and fitting amplitude, background, and the endpoint energy. Compare with the time distribution in fig. \ref{fig:mnu} optimized with respect to the standard deviation of the neutrino mass.}
	\label{fig:q}
\end{figure}

\section{Other applications}

The method can also be effectively used to show what energy region of the measured spectrum affects mostly a standard deviation of a selected parameter. As an example, we focused on the endpoint of beta spectrum. Since the beta spectrum endpoint and the neutrino mass are strongly correlated \cite{katrin_loi, katrin_des}, we fixed the neutrino mass to $0\,$eV/$c^2$. Then, the optimal time distribution minimizing the standard deviation of the endpoint was derived. It is shown in fig. \ref{fig:q}.

The optimal distribution of measurement time was found following the method described in the above paragraphs using the same standard distribution of the measurement points as in the previous example. Here, it was the standard deviation of the tritium endpoint we were focused on, not the standard deviation of the neutrino mass. Again, starting with a uniform time distribution 10 iterations were performed. The method sets longer measurement times in the measured points that are sensitive to the beta spectrum endpoint.

We would like to note that the same method can be used to suppress systematics, i.e., to find a region of the spectrum, which is the most sensitive to systematics, and then to exclude the region from the set of the measurement points.

\section{Discussion}

A method to distribute the measurement time in single channel measurements into the measurement points to minimize the standard deviation of the selected fit parameter was offered and demonstrated on two examples. It worked the desired way. Even more, the method proved useful to show what energy region of a measured spectrum is sensitive to a selected parameter.

In place of summary, we would like to emphasize, that the method is mathematical. And so are its results. Although the method assumed a fixed distribution of measurement points, this assumption should be reconsidered if necessary. E.g., if the optimal distribution of measurement time aggregates most of the measurement time into few measurement points, then a more dense distribution of points in the particular measurement region is appropriate. This actually happened in the KATRIN example. The case of measurement points with negligible measurement time can be treated in a similar way. An extraordinary care should be paid to a case when regions sensitive to the selected parameter overlap regions sensitive to systematics.

\begin{ack}
This work was partly supported by the Grant Agency of the Czech Republic under contract No.\,202/06/0002, and by the Ministry of Education, Youth, and Sports of the Czech Republic under contracts No.\,LA 318, and LC 07050.
\end{ack}

\end{document}